\documentclass[10pt,twocolumn,letterpaper]{article}

\usepackage{wifs}
\usepackage{times}
\usepackage{epsfig}
\usepackage{graphicx}
\usepackage{amsmath}
\usepackage{amssymb}
\usepackage{multirow}
\usepackage{comment}
\usepackage{fancyhdr}

\usepackage[pagebackref=true,breaklinks=true,letterpaper=true,colorlinks,bookmarks=false]{hyperref}

\wifsfinalcopy 

\def\wifsPaperID{86} 

\ifwifsfinal\fi
\begin{document}


\newcommand{\E}{\mathbb E}
\newcommand{\argmax}{\mathop{\rm max}\limits}
\newcommand{\argmin}{\mathop{\rm min}\limits}

\title{Transforming acoustic characteristics to deceive playback spoofing countermeasures of speaker verification systems}

\author{Fuming Fang$^1$, Junichi Yamagishi$^{1,2}$, Isao Echizen$^1$, Md Sahidullah$^3$, Tomi Kinnunen$^4$ \\
$^1$National Institute of Informatics, Japan$~~$ 
$^2$University of Edinburgh, UK \\
$^3$Inria, France$~~$ 
$^4$University of Eastern Finland, Finland \\
{\tt\small \{fang, jyamagis, iechizen\}@nii.ac.jp, md.sahidullah@inria.fr, tkinnu@cs.uef.fi}
}

\maketitle

\begin{abstract}
Automatic speaker verification (ASV) systems use a playback detector to filter out playback attacks and ensure verification reliability. Since current playback detection models are almost always trained using genuine and played-back speech, it may be possible to degrade their performance by transforming the acoustic characteristics of the played-back speech close to that of the genuine speech. One way to do this is to enhance speech ``stolen'' from the target speaker before playback. We tested the effectiveness of a playback attack using this method by using the speech enhancement generative adversarial network to transform acoustic characteristics.
Experimental results showed that use of this ``enhanced stolen speech'' method significantly increases the equal error rates for the baseline used in the ASVspoof 2017 challenge and for a light convolutional neural network-based method.
The results also showed that its use degrades the performance of a Gaussian mixture model-universal background model-based ASV system. This type of attack is thus an urgent problem needing to be solved. 
\end{abstract}

\thispagestyle{fancy}
\fancyhf{}
\renewcommand{\headrulewidth}{0pt}
\chead{\small Accepted to be Published in IEEE International Workshop on Information Forensics and Security (WIFS) 2018, Hong Kong, China}
\pagestyle{empty}
\lfoot{\small \copyright 2018 IEEE. Personal use of this material is permitted. Permission from IEEE must be obtained for all other uses, in any current or future media, including reprinting/republishing this material for advertising or promotional purposes, creating new collective works, for resale or redistribution to servers or lists, or reuse of any copyrighted component of this work in other works.}

\vspace{-1mm}
\section{Introduction}
\vspace{-1mm}
Automatic speaker verification (ASV)~\cite{hansen2015speaker}, a kind of biometrics authentication technology, identifies a person from a segment of speech. ASV systems typically fall into two types: text-independent and text-dependent, where the latter requests a client to speak a given phrase. Due to the convenience of ASV, it is being used in more and more applications, such as ones used in call centers and by mobile devices. However, ASV is vulnerable to several kinds of spoofing attacks (also known as presentation attacks~\cite{isopad}), so ASV systems need a spoofing countermeasure (CM) (also known as presentation attack detection~\cite{isopad}). Such attacks aim to mimic the target speaker mainly by using synthesized speech~\cite{wu2015asvspoof}, converted speech~\cite{wu2015asvspoof}, or playback speech~\cite{EURECOM+5235, EURECOM+5504}. 
Among them, playback speech-based attacks are relatively easy to mount since an attacker who has no special knowledge can make them~\cite{wu2015spoofing}.
Once an attacker has collected/stolen a voice sample for the target speaker, he/she can simply play it back to an ASV system or concatenate segments of the sample to form a new utterance. Threats from this kind of attack have been confirmed by several studies~\cite{EURECOM+5235, EURECOM+5504, alegre2014re, galka2015playback, wu2014a}. Here we focus on playback spoofing attacks and relevant CMs.

Four main types of CMs have been developed to protect against playback spoofing attacks. \emph{One type} utilizes a text-dependent ASV system and randomly prompts for a pass-phrase~\cite{EURECOM+4934, Zeinali2018}, making it difficult to mount playback attacks using phrase-fixed speech. However, it is possible to form an arbitrary utterance to spoof this type of CM if the attacker has sufficient speech data for the target speaker. \emph{The second type} is based on rules describing the characteristics of genuine speech (recorded from a person). For example, Mochizuki et al.~\cite{Mochizuki2018} distinguished genuine speech by detecting pop-noise from certain phonemes. An intractable problem related to this type of CM is that it is difficult to design suitable rules and implement them. \emph{The third type} utilizes audio fingerprinting to check whether an incoming recording is similar to previously authenticated utterances that were automatically saved in the ASV system. Rodriguez et al.~\cite{Gonzalez2018an} developed such a system: if the similarity score was higher than a threshold, the recording was treated as a playback attack. A disadvantage of this type of CM is that it is sensitive to noise.
In contrast, \emph{the fourth type} compares the differences between genuine speech and playback speech. This type mainly utilizes a machine learning algorithm to learn the differences. An example is Wang et al.'s~\cite{wang2016efficient} use of a support vector machine~\cite{hearst1998support} to learn the difference in Mel-frequency cepstral coefficient (MFCC)-based acoustic features. 

More methods of the fourth type were presented at the second Automatic Speaker Verification Spoofing and Countermeasures Challenge (ASVspoof 2017), in which a common database was used to assess the participants' CMs.
The database consists of two parts. One part contains genuine speech taken from the RedDots corpus~\cite{lee2015reddots}, which was designed for speaker verification. The other part contains recordings of the genuine speech made in various environments. For these data, the baseline~\cite{EURECOM+5235} with a constant Q cepstral coefficient (CQCC)~\cite{todisco2016new} feature and a Gaussian mixture model (GMM) classifier had an equal error rate (EER) of 30.60\%. A deep learning-based method had an EER of 6.73\% ~\cite{lavrentyeva2017audio}, which was the best performance achieved at ASVspoof 2017~\cite{EURECOM+5235}.

These mainstream CMs of the fourth type are also problematic: they are based on the assumption that the attackers do not have special knowledge.
Moreover, this type of CM algorithms only learn the difference from a given dataset and perhaps do not work well if the acoustic characteristics of the playback speech is transformed close to that of the genuine one. To confirm this hypothesis, we tested the effectiveness of a playback attack using speech ``stolen'' from the target speaker and enhanced before mounting the attack. This enhancement should remove the distortions in the stolen speech caused by the recording device and environmental noise so that they do not affect the re-recorded speech.

\begin{figure}[t]
\begin{center}
    \includegraphics[width=0.8\linewidth]{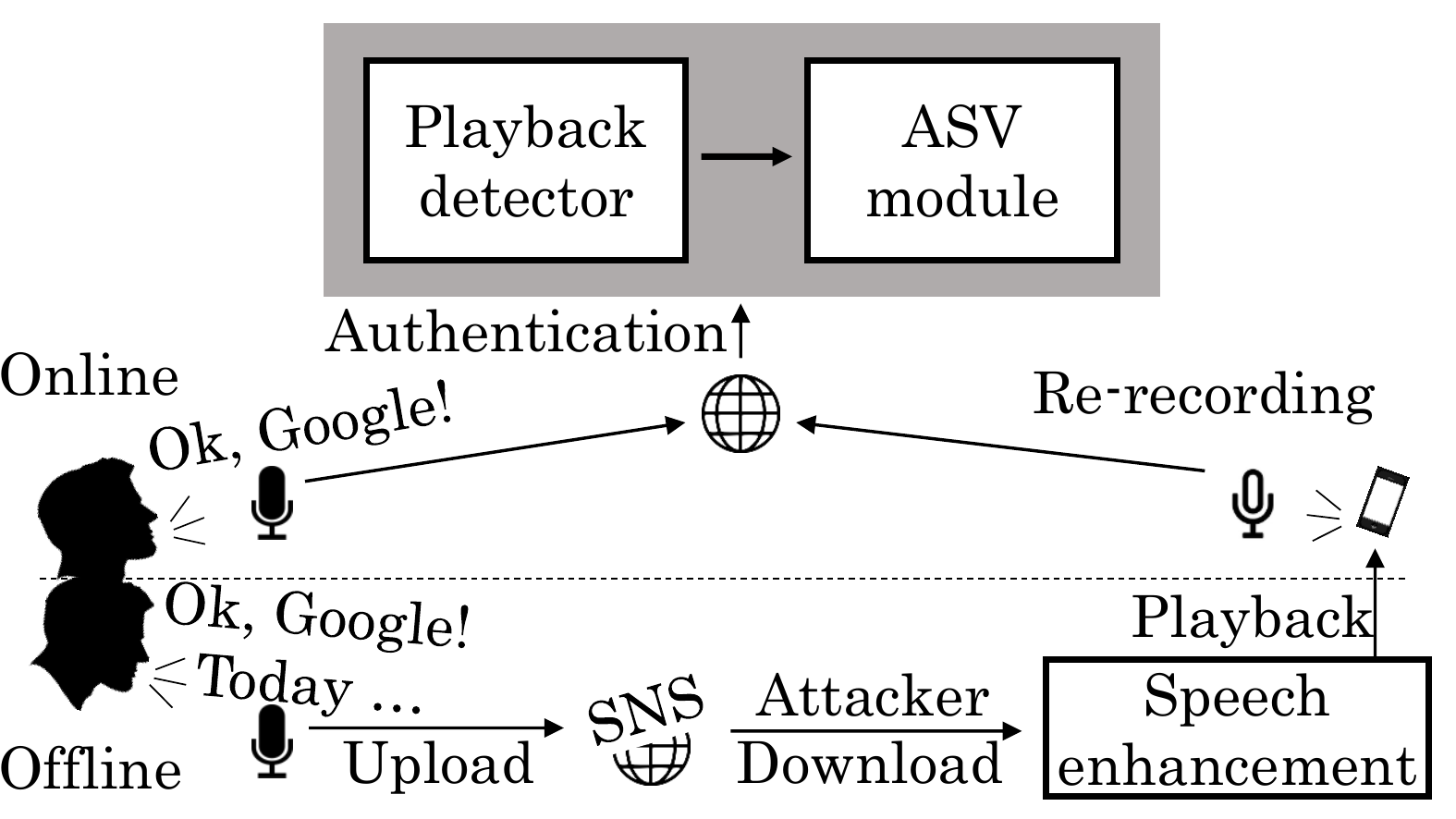}
\end{center}
\vspace{-4mm}
\caption{Playback spoofing attack using enhanced stolen speech method under ASVspoof 2017 scenario. Without speech enhancement, attack is the same as a conventional playback attack.}
\vspace{-4mm}
\label{fig:proposal}
\end{figure}

We evaluated the effectiveness of a playback attack using this “enhanced stolen speech” method against a text-dependent ASV system. 
We used the ASVspoof 2017 scenario (Figure~\ref{fig:proposal}) in which the attacker is assumed to obtain from somewhere uncompressed speech for the target speaker containing the phrase used for authentication, e.g., by downloading from the web, hacking a device used by the target speaker, and talking to and surreptitiously recording the target speaker.
The speech enhancement generative adversarial network (SEGAN)~\cite{pascual2017segan} was used to transform the acoustic characteristics of the obtained speech close to that of the genuine speech.
We also investigated the effect of different types of playback loudspeakers and re-recording devices. The results showed that it is possible to fool playback spoofing CMs by transforming the acoustic characteristic of the playback speech close to that of the genuine speech.

\vspace{-2mm}
\section{Related work}
\vspace{-2mm}
\label{sec:related_work}
Pioneering work on playback attacks was reported by Lindberg and Blomberg in 1999~\cite{lindberg1999vulnerability}. They pre-recorded the numbers one to ten of two speakers and then concatenated various combinations of them to attack a hidden Markov model (HMM)~\cite{rabiner1989tutorial}-based text-dependent ASV system. They demonstrated considerable increase in both the EER and false acceptance rate (FAR) compared to verification without attacks. More recently, Ergunay et al. investigated the effect of playback attacks against ASV systems and also achieved a large increase in FAR~\cite{ergunay2015vulnerability}. 
Compared to these conventional playback attacks, our method further degrades the performance of playback spoofing CMs by enhancing the speech.

There are a few attack methods similar to our enhanced stolen speech method. Demiroglu et al. improved the naturalness of synthesized and converted speech before attacking a phase-based synthetic speech detector and an ASV system~\cite{demiroglu2017postprocessing}. The synthesized and converted speech signals were firstly analyzed frame by frame, and each frame was replaced with one containing the most similar natural speech selected from a dataset. A complex cepstrum vocoder was used to re-synthesize these frames so as to improve speech naturalness. Finally, the speech was directly fed into an ASV system. They reported that their method fooled four out of nine detectors. Our method can be thought of as an extension of their method as it further transforms synthesized speech close to natural speech.

Nguyen et al. reported an attack method that transforms computer-generated (CG) images into natural images before feeding them into a facial authentication system~\cite{nguyen2018transformation}. The transformation model is trained using a generative adversarial network (GAN)~\cite{goodfellow2014generative}. The GAN discriminator, which mimics a spoofing detector, is used to distinguish CG/natural images. The discriminator is pre-trained and fixed during training of the transformation model. In contrast, we treat the authentication system as a black box, and anything regarding playback spoofing CMs and ASV systems is unknown.

\vspace{-2mm}
\section{Playback detectors and ASV system}
\label{sec:asv}
\vspace{-2mm}
Two playback spoofing CMs and a classical Gaussian mixture model with universal background model (GMM-UBM)~\cite{reynolds2000speaker}-based ASV system were used to evaluate the effectiveness of our enhanced stolen speech attack method. The two CMs were the baseline and the core method of the system with the best performance (i.e., a light convolutional neural network) of ASVspoof 2017. 

\vspace{-1mm}
\subsection{Baseline of ASVspoof 2017}
\vspace{-1mm}
The baseline of ASVspoof 2017 consists of a CQCC front-end and a GMM back-end. We refer to this method as ``CQCC-GMM CM''. The CQCC is an acoustic feature extracted from an audio signal. CQCC extraction is performed using constant Q transform (CQT) instead of the classical short-time Fourier transformation (STFT). STFT suffers from fixed frequency resolution and fixed temporal resolution whereas CQT exhibits higher frequency resolution at lower frequencies and higher temporal resolution at higher frequencies. An audio signal is usually represented by a sequence of CQCC feature vectors.

A two-class GMM-based classifier is used for genuine/playback speech detection. One GMM is trained using genuine speech while the other one is trained using playback speech. Input to the GMMs is CQCC-based acoustic feature vectors, and the expectation maximization (EM)~\cite{dempster1977maximum} algorithm is used for training. During prediction, the feature vectors of an audio signal are independently input into the two models, and then the joint log-likelihood for both models is calculated. Finally, the log-likelihood ratio of the genuine and playback model results is compared with a threshold to determine genuine/playback speech.

\vspace{-1mm}
\subsection{Core method of best system of ASVspoof 2017}
\vspace{-1mm}
The best system of ASVspoof 2017 was a fusion of three sub-systems: a support vector machine with $i$-vector features~\cite{dehak2011front}, a convolutional neural network (CNN) with a recurrent neural network (RNN), and a light CNN (LCNN). The LCNN was used as the core method, which achieved an EER of 7.37\%. This performance was very close to that of the fused system (6.73\%). We therefore used an ``LCNN CM'' to evaluate our enhanced stolen speech attack method. 

The LCNN consists of five convolution layers, four network-in-network (NIN)~\cite{lin2013network} layers, ten max-feature-map (MFM) layers, five max-pooling layers, and two fully connected layers. Each MFM layer acts as a maxout activation function~\cite{goodfellow2013maxout} that splits the CNN feature maps into two groups and then performs element-wise maximization to select features. 
The LCNN input is a spectrum with a fixed size of $864 \times 400$, which is obtained by performing STFT with 1728 bins and concatenating 400 frames. Dropout~\cite{srivastava2014dropout} is applied after the first fully connected layer. The final output layer, with a softmax activation function, is used to discriminate genuine/playback speech. This is described in more detail elsewhere~\cite{lavrentyeva2017audio}.

The silence parts of the audio signal are removed, and then STFT is performed using a window length of 25 ms with a shift size of 10 ms.
If a signal is shorter than 4 seconds, its content is repeated to match the length. For a longer signal, its content is repeated to match multiples of 4 seconds, and the output probabilities are averaged. 

\vspace{-2mm}
\subsection{GMM-UBM-based ASV system}
\vspace{-2mm}
We use a GMM-UBM-based system for speaker verification. Even though it is a classic ASV method, GMM-UBM provides competitive performance on short-duration, text-dependent ASV tasks~\cite{delgado2016further}. The speaker models are created by maximum a posteriori adaptation from a UBM trained with a large amount of speech data from different speakers. Text-dependent speaker models are separately created for different passphrases following the guidelines for conducting experiments with the RedDots corpus.
The recognition score is the likelihood ratio between the results of the target speaker model and those of the UBM.

\vspace{-1mm}
\section{Speech enhancement}
\label{sec:segan}
\vspace{-1mm}
SEGAN is a data-driven speech enhancement method that constructs a mapping from a noisy waveform to a clean waveform with the help of supervised training. More specifically, SEGAN leverages the power of the GAN composed of two adversarial networks, a discriminator $D$ and a generator $G$.
The discriminator predicts the probability that the input is from real data rather than from fake data generated by $G$. The generator learns a mapping function from a prior noise distribution $p_{\mathrm{noise}}({\bf z})$ to the distribution of the real data $p_{\mathrm{data}}({\bf y})$ to fool the discriminator. If the noise distribution is conditioned by ${\bf x}$ drawn from playback speech and ${\bf y}$ drawn from genuine speech, the output $\hat{\bf y}$ is genuine-like speech.

The objective function for training SEGAN is formulated as 
\vspace{-3mm}
\begin{align}
\argmin_{D} \mathcal{L}(D) &=\frac{1}{2}\E_{{\bf y}\sim p_{\mathrm{data}}({\bf y})}[(D({\bf y}, {\bf x}) - 1)^2] \nonumber \\
&+\frac{1}{2}\E_{{\bf z}\sim p_{\mathrm{noise}}({\bf z})}[D(\hat{\bf y}, {\bf x})^2] \nonumber \\
\argmin_{G} \mathcal{L}(G) &=\E_{{\bf z}\sim p_{\mathrm{noise}}({\bf z})}[(D(\hat{{\bf y}}, {\bf x}) - 1)^2] \nonumber \\
&+\lambda \cdot \parallel \hat{\bf y} - {\bf y} \parallel_1,
\label{eq:segan2}
\end{align}
where $\hat{\bf y} = G({\bf z}, {\bf x})$ is enhanced (or generated) speech and $\E$ means expectation. $L_1$ norm, $\parallel \cdot \parallel_1$, is used to measure the distance between the real and enhanced speech. The discriminator and generator are alternately trained by performing a min-max game. 

\begin{figure}[t]
\begin{center}
    \includegraphics[width=0.6\linewidth]{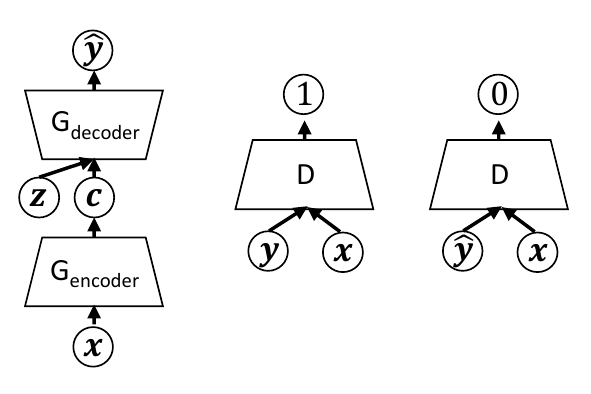}
\end{center}
\vspace{-6mm}
\caption{Architecture of SEGAN. Input of discriminator is $({\bf y}, {\bf x})$ or $(\hat{\bf y}, {\bf x})$; the former should be classified as real data while the latter should be classified as fake data; but the latter should be treated as real data when updating the generator parameters, so adversarial training is performed.}
\vspace{-4mm}
\label{fig:segan}
\end{figure}
Figure~\ref{fig:segan} shows the architecture of SEGAN. It is an end-to-end model, so both the input and output of the generator are raw waveforms. The input of the discriminator is also a raw waveform combining $({\bf y}, {\bf x})$ or $(\hat{\bf y}, {\bf x})$. The generator has an encoder-decoder structure. The encoder part is composed of 11 stacked 1-D CNN layers with a filter width of 31 and a stride of two. The decoder part is a mirror structure of the encoder part, and the corresponding layers between them are connected by a skip path.
The dimension of noise ${\bf z}$ is the same as that of encoder output ${\bf c}$ and is drawn from a normal distribution. They are concatenated and input to the decoder. The architecture of the discriminator is the same as that of the encoder part of the generator except that virtual batch-normalization~\cite{salimans2016improved} is performed in hidden layers.

\vspace{-1mm}
\section{Database}
\label{sec:database}
\vspace{-2mm}
 In this section, we describe the speech data used for training the spoofing detectors, ASV, and SEGAN. We also describe the test data used for authentic as well as illegitimate access.

\vspace{-1mm}
\subsection{Training data for playback spoofing CM}
\vspace{-2mm}
Both the CQCC-GMM and the LCNN CM models were trained using the ASVspoof 2017 database (version 2), which was derived from the RedDots corpus. The genuine speech data in the database was taken from the RedDots corpus, and the playback speech data was recorded by playing the genuine speech data in various environments (including quiet and noisy places) using various recording devices and speakers with various qualities. The sampling rate was 16 kHz for both the genuine and playback speech. The database was further split into three datasets: training, development, and evaluation. Each of the datasets contained both genuine and playback speech. The GMM of the CQCC-GMM CM was trained using the training and development datasets. For the LCNN, the training dataset was used to estimate the model parameters and the development dataset was used to monitor the training process.

\vspace{-1mm}
\subsection{Training data for ASV system}
\vspace{-2mm}
We used TIMIT and RSR2015 (background subset~\cite{larcher2014text}) corpora for training the UBM for the GMM-UBM-based ASV system. Only male speakers were used as the ASVspoof 2017 database was created from the male subset of the RedDots corpus. In total, we used 17,850 speech utterances from 488 speakers for UBM training. Each target speaker model was created with speech utterances from three different sessions for a fixed passphrase.

\vspace{-1mm}
\subsection{Training data for SEGAN}
\vspace{-1mm}
We used a high-quality database and two low-quality databases distorted by recording devices or environment noises to train SEGAN. The high-quality database was the voice cloning toolkit (VCTK) corpus~\cite{vctk}. This corpus contains data recorded in a hemi-anechoic chamber by 109 native English speakers, but we used data for only 28 speakers. One of the low-quality databases was a device-recorded VCTK (DR-VCTK) corpus~\cite{drvctk} and the other one was a noisy VCTK (N-VCTK) corpus~\cite{nvctk}. The DR-VCTK was created by playing the high-quality speech of the 28 speakers in office environments and recording it using relatively inexpensive consumer devices. The N-VCTK was created by adding noise to the high-quality speech of the 28 speakers.
The sampling rate of these databases was 48 kHz with downsampling to 16 kHz. Two types of SEGAN were trained. One was trained using DR-VCTK and VCTK. The other was trained using N-VCTK and VCTK.

\vspace{-1mm}
\subsection{Authentication and spoofing data}
\vspace{-1mm}
We equally split the genuine speech of the evaluation dataset in the ASVspoof 2017 database into two sub-datasets. One was used as authentication speech, and the other was used as ``stolen speech.'' We enhanced the stolen speech, played it using four types of portable loudspeakers, and re-recorded it using six types of recording devices in an office room. The four loudspeakers were a high-quality speaker (BOSE Soundlink), a medium-quality speaker (SONY SRS-BTS50), a low-quality speaker (audio-technica AT-SP92), and an iPhone 6s speaker. The six recording devices were a high-quality condenser microphone (Apogee MiC 96k), a directional microphone (Sony ECM--673), a low-quality microphone (Snowball iCE), a MacBook microphone, an iPad microphone, and an iPhone 6s microphone. These devices were placed at around 30 to 50 cm from the loudspeaker. The sampling rate for the re-recording was 16 kHz. 
According to the used loudspeaker, four playback and re-recording sessions were performed.

\begin{table*}
\caption{EERs for CQCC-GMM CM. Bold means largest degradation.}
\label{tbl:res_cqcc_gmm}
\begin{center}
\begin{tabular}{|c|c|cccccc|c|}
\hline
Loudspeaker & Enhancement & Directional&High-quality&Low-quality&Mac & \multirow{2}{*}{iPad} & iPhone & \multirow{2}{*}{Average} \\
used for replay & training data &  microphone &  microphone &  microphone & book&&6s& \\
\hline\hline
\multirow{3}{*}{High quality} & $-$ & 15.65 &	8.83 &	20.87 &	9.98 &	7.21 &	49.92 & 18.74  \\
&DR-VCTK & 28.42 &	18.10 &	29.67 &	14.96 &	8.59 &	50.00  & 24.96  \\
&N-VCTK & 35.61 &	{\bf 23.18} &	33.59 &	16.49 &	9.17 &	50.00  & 28.01 \\
\hline
\multirow{3}{*}{Medium quality}& $-$ & 9.35 &	11.96 &	8.78 &	10.78 &	6.54 &	49.13 & 16.09 \\
&DR-VCTK & 15.71 &	20.16 &	15.76 &	15.27 &	7.43 &	49.92 & 20.71 \\
&N-VCTK & 22.56 &	25.62 &	22.03 &	15.61 &	8.36 &	49.92  & 24.02 \\
\hline
\multirow{3}{*}{Low quality}  & $-$ & 11.83 &	8.98 &	10.28 &	8.34 &	6.14 &	49.92 & 15.92  \\
&DR-VCTK & 20.07 &	16.29 &	19.77 &	10.32 &	6.96 &	49.96 & 20.56 \\
&N-VCTK & 26.87 &	22.35 &	24.44 &	10.78 &	7.35 &	49.92 &	23.62  \\
\hline
\multirow{3}{*}{iPhone 6s} & $-$ & 16.28 &	16.54 &	19.83 &	7.19 &	6.40 &	49.53 &	19.30 \\
&DR-VCTK & 30.45 &	31.19 &	30.50 &	10.93 &	7.14 &	49.92 &	26.69 \\ 
&N-VCTK & 24.25 &	24.26 &	26.94 &	9.83 &	7.28 &	49.88 &	23.74 \\
\hline
\end{tabular}
\end{center}
\vspace{-4mm}
\end{table*}
\begin{table*}
\caption{EERs for LCNN CM. Bold means largest degradation.}
\label{tbl:res_lcnn}
\begin{center}
\begin{tabular}{|c|c|cccccc|c|}
\hline
Loudspeaker & Enhancement & Directional&High-quality&Low-quality&Mac & \multirow{2}{*}{iPad} & iPhone & \multirow{2}{*}{Average} \\
used for replay & training data &  microphone &  microphone &  microphone & book&&6s& \\
\hline\hline
\multirow{3}{*}{High quality} & $-$ & 11.19 &	8.00 &	16.14 &	7.71 &	12.95 &	25.04 &	13.51  \\
&DR-VCTK & 12.35 &	9.12 &	18.14 &	8.59 &	13.55 &	25.74 &	14.58  \\
&N-VCTK & 13.48 &	10.43 &	19.74 &	8.85 &	13.83 &	25.29 &	15.27  \\
\hline
\multirow{3}{*}{Medium quality}& $-$ & 8.78 &	9.98 &	5.92 &	5.47 &	7.09 &	25.25 &	10.42  \\
&DR-VCTK & 9.57 &	11.22 &	6.56 &	6.85 &	8.79 &	27.25 &	11.71  \\
&N-VCTK &  10.31 &	12.22 &	7.96 &	7.23 &	9.56 &	27.12 &	12.40 \\
\hline
\multirow{3}{*}{Low quality}  & $-$ & 7.25 &	6.06 &	5.31 &	9.52 &	7.80 &	16.29 &	8.71  \\
&DR-VCTK & 8.44 &	7.10 &	6.07 &	10.08 &	8.76 &	17.05 &	9.58  \\
&Noisy VCTK & 10.23 &	{\bf 8.95} &	7.52 &	10.30 &	9.38 &	17.09 &	10.58  \\
\hline
\multirow{3}{*}{iPhone 6s} & $-$ & 11.11 &	11.56 &	10.47 &	4.40 &	9.17 &	17.65 &	10.73  \\
&DR-VCTK & 13.25 &	14.97 &	13.62 &	5.23 &	11.12 &	18.26 &	12.74  \\ 
&N-VCTK & 11.70 &	12.33 &	11.21 &	4.54 &	10.07 &	18.07 &	11.32  \\
\hline
\end{tabular}
\end{center}
\vspace{-8mm}
\end{table*}
\begin{table*}
\caption{Values of t-DCF obtained from combination of CQCC-GMM CM and ASV scores. Bold means largest degradation.}
\label{tbl:tdcf_cqccgmm}
\begin{center}
\begin{tabular}{|c|c|cccccc|c|}
\hline
Loudspeaker & Enhancement & Directional&High-quality&Low-quality&Mac & \multirow{2}{*}{iPad} & iPhone & \multirow{2}{*}{Average} \\
used for replay & training data &  microphone &  microphone &  microphone & book&&6s& \\
\hline\hline
\multirow{3}{*}{High quality} & $-$ & 0.9361 &	0.9276 &	0.9392 &	0.9136 &	0.9118 &	0.9426 &	0.9285   \\
&DR-VCTK & 0.9412 &	0.9363 &	0.9410 &	0.9258 &	0.9210 &	0.9426 &	0.9347  \\
&N-VCTK &  0.9403 &	0.9373 &	0.9402 &	0.9277 &	0.9211 &	0.9431 &	0.9350   \\
\hline
\multirow{3}{*}{Medium quality}& $-$ &  0.9156 &	0.9351 &	0.9211 &	0.9222 &	0.9039 &	0.9425 &	0.9234   \\
&DR-VCTK &  0.9313 &	0.9386 &	0.9348 &	0.9311 &	0.9143 &	0.9428 &	0.9322   \\
&N-VCTK &  0.9339 &	0.9377 &	0.9362 &	0.9308 &	0.9186 &	0.9429 &	0.9334   \\
\hline
\multirow{3}{*}{Low quality}  & $-$ &  0.9225 &	0.9177 &	0.9248 &	0.9109 &	0.9020 &	0.9415 &	0.9199   \\
&DR-VCTK &  0.9339 &	0.9314 &	0.9362 &	0.9220 &	0.9076 &	0.9428 &	0.9290   \\
&N-VCTK &  0.9353 &	0.9342 &	0.9363 &	0.9223 &	0.9105 &	0.9425 &	0.9302   \\
\hline
\multirow{3}{*}{iPhone 6s} & $-$ & 0.9354 &	0.9358 &	0.9379 &	0.9085 &	0.9006 &	0.9384 &	0.9261    \\
&DR-VCTK &  0.9380 &	0.9379 &	0.9388 &	{\bf 0.9290} &	0.9127 &	0.9381 &	0.9324   \\ 
&N-VCTK &  0.9387 &	0.9391 &	0.9383 &	0.9247 &	0.9100 &	0.9388 &	0.9316   \\
\hline
\end{tabular}
\end{center}
\vspace{-4mm}
\end{table*}
\begin{table*}
\caption{Values of t-DCF obtained from combination of LCNN CM and ASV scores. Bold means largest degradation.}
\label{tbl:tdcf_lcnn}
\begin{center}
\begin{tabular}{|c|c|cccccc|c|}
\hline
Loudspeaker & Enhancement & Directional&High-quality&Low-quality&Mac & \multirow{2}{*}{iPad} & iPhone & \multirow{2}{*}{Average} \\
used for replay & training data &  microphone &  microphone &  microphone & book&&6s& \\
\hline\hline
\multirow{3}{*}{High quality} & $-$ & 0.9239 &	0.9073 &	0.9436 &	0.9063 &	0.9338 &	0.9656 &	0.9301 \\
&DR-VCTK & 0.9303 &	0.9135 &	0.9494 &	0.9098 &	0.9385 &	0.9664 &	0.9347  \\
&N-VCTK &   0.9346 &	0.9186 &	0.9522 &	0.9105 &	0.9390 &	0.9658 &	0.9368  \\
\hline
\multirow{3}{*}{Medium quality}& $-$ & 0.9105 &	0.9173 &	0.8990 &	0.8978 &	0.9038 &	0.9657 &	0.9157    \\
&DR-VCTK &  0.9159 &	0.9266 &	0.9011 &	0.9021 &	0.9107 &	0.9673 &	0.9206   \\
&N-VCTK &  0.9202 &	0.9293 &	0.9077 &	0.9035 &	0.9143 &	0.9675 &	0.9238   \\
\hline
\multirow{3}{*}{Low quality}  & $-$ &  0.9043 &	0.8992 &	0.8972 &	0.9122 &	0.9070 &	0.9584 &	0.9131   \\
&DR-VCTK & 0.9105 &	0.9029 &	0.8999 &	0.9156 &	0.9125 &	0.9605 &	0.9170    \\
&N-VCTK &  0.9183 &	0.9108 &	0.9051 &	0.9156 &	0.9141 &	0.9593 &	0.9205   \\
\hline
\multirow{3}{*}{iPhone 6s} & $-$ & 0.9208 &	0.9238 &	0.9173 &	0.8952 &	0.9120 &	0.9530 &	0.9204    \\
&DR-VCTK &  0.9344 &	0.9396 &	{\bf 0.9345} &	0.8970 &	0.9222 &	0.9540 &	0.9303   \\ 
&N-VCTK &  0.9234 &	0.9274 &	0.9214 &	0.8960 &	0.9154 &	0.9559 &	0.9233   \\
\hline
\end{tabular}
\end{center}
\vspace{-8mm}
\end{table*}

\vspace{-3mm}
\section{Experimental setup}
\label{sec:setup}
\vspace{-1mm}
We evaluated 1) how our enhanced stolen speech method affects the performance of playback spoofing CMs and 2) how the enhanced speech effects an ASV system. In order to compare the difference between our method and conventional playback attacks, a paired two-tailed $t$-test was used.

\vspace{-1mm}
\subsection{Setup for playback spoofing CMs}
\vspace{-1mm}
Settings of the CQCC-GMM CM were the same as baseline of ASVspoof 2017 and the source code is available at~\cite{asvspoof2017baseline}.
CQCC had 29 dimension and 0-th order cepstral coefficient was further used. Their first and second derivatives were finally used as features (90 dimensions in total). The GMM of the CQCC-GMM CM had 512 components. 

The weights of the LCNN were initialized using the Xavier method~\cite{glorot2010understanding}. The dropout rate was set to 0.5. Adam optimization~\cite{kingma2014adam} with momentum of 0.5 was used. The initial learning rate was 0.0001; it was reduced by 0.9 if the classification accuracy of the development dataset decreased after each epoch. There were nine epochs, and the mini-batch size was 64. The LCNN was implemented using the TensorFlow framework~\cite{abadi2016tensorflow} and is available at~\cite{lcnncode}.

To assess the performance of both spoofing CMs, we use EER, which reflects the ability of the CM to discriminate genuine speech samples from playback attacks. 

\vspace{-1mm}
\subsection{Setup for ASV system}
\vspace{-1mm}
Our ASV system used MFCC-based acoustic feature extracted from a 20~ms short-term window with a 10~ms shift using 20 filters. We computed 19 MFCCs after discarding the energy coefficients. The MFCCs were further processed with RASTA filtering to suppress convolutive mismatch. The delta and double-delta coefficients were computed for a context of three frames and then augmented with static MFCCs to create a 57-dimensional feature vector. Finally, cepstral mean and variance normalization (CMVN) was performed after discarding the non-speech frames with an energy-based voice activity detector. We trained the gender-dependent UBM with 512 mixture components. The speaker models were created by adapting only the centers of UBM with a relevance factor of three.

\vspace{-1mm}
Even though ASV spoofing evaluations have focused on standalone CM assessment, the performance of a tandem (combined) system is important for real-world deployment. Both CM and ASV can result in target speaker misses and false acceptances of impostors (either non-targets or spoofs). We therefore adopted a recently proposed \emph{tandem detection cost function} (t-DCF) metric~\cite{Kinnunen2018tDCFAD} for evaluating the combination of two systems in a Bayes risk framework. The t-DCF is given by $C_{\mathrm{miss}}^{\mathrm{asv}} \cdot \pi_{\mathrm{tar}} \cdot P_a + C_{\mathrm{fa}}^{\mathrm{asv}} \cdot \pi_{\mathrm{non}} \cdot P_b + C_{\mathrm{fa}}^{\mathrm{cm}}  \cdot \pi_{\mathrm{spoof}} \cdot P_c + C_{\mathrm{miss}}^{\mathrm{cm}} \cdot \pi_{\mathrm{tar}} \cdot P_d$, where $C_{\mathrm{miss}}^{\mathrm{asv}}$ = $1$, $C_{\mathrm{fa}}^{\mathrm{asv}}$ = $10$, $C_{\mathrm{fa}}^{\mathrm{cm}}$ = $10$, and $C_{\mathrm{miss}}^{\mathrm{cm}}$ = $1$ are unit costs related to the misses and false alarms of the two systems; $\pi_{\mathrm{spoof}}$ = $0.0100$, $\pi_{\mathrm{non}}$ = $0.0099$, and $\pi_{\mathrm{tar}}$ = $0.9801$ were the prior probabilities of the targets, non-targets, and spoofs, respectively; and $P_a$, $P_b$, $P_c$, and $P_d$ are the error rates of four possible errors originating from the joint actions of the CM and ASV systems. The reported t-DCF values are minimum t-DCF values with a fixed ASV system.
The higher the value, the less usable the combined (ASV and CM) system.

\vspace{-2mm}
\subsection{Setup for SEGAN}
\vspace{-1mm}
Similar to previous work~\cite{pascual2017segan, lorenzo2018can}, we extracted chunks of waveforms by using a sliding window of $2^{14}$ samples at every $2^{13}$ samples (i.e., 50\% overlap). During testing, we concatenated the results at the end of the stream without overlap. The learning rate, mini-batch size, and epoch size were set to 0.0002, 100, and 120, respectively. The $\lambda$ in Equation~\ref{eq:segan2} was set to 100.
We used source code for improved SEGAN~\cite{improvedsegan}.

\vspace{-2mm}
\section{Results}
\label{sec:results}
\vspace{-1mm}
Tables~\ref{tbl:res_cqcc_gmm} and \ref{tbl:res_lcnn} show the EERs for the CQCC-GMM CM and the LCNN CM, respectively. Playback spoofing attacks using our enhanced stolen speech method had significantly higher EERs for both CMs compared to those of conventional playback attacks (without enhancement). One reason could be that the signal-to-noise ratio was higher after speech enhancement, resulting in the playing of cleaner speech. Use of the high-quality speaker with the high-quality microphone and use of the low-quality speaker with the high-quality microphone when N-VCTK was used to train SEGAN resulted in the largest performance degradation for the two CMs. The increases in EER were 2.6 and 1.5 times, respectively.

\vspace{-2mm}
As expected, use of the high-quality speaker resulted in higher EERs because it generated more natural speech. It is interesting that the results for the iPhone 6s speaker were similar to those for the high-quality speaker. While a wide range of EERs were obtained for the recording devices, use of the high-quality microphone did not result in significantly higher EERs. The CQCC-GMM CM could not distinguish the playback speech re-recorded using the iPhone 6s.
This was because features were not normalized and channel distortions greatly degraded its performance~\cite{EURECOM+5504}.
Enhancement based on the N-VCTK was mightier than that based on the DR-VCTK in most cases. This might be because distortion due to environmental noise has a greater effect than that due to the recording devices.

Tables~\ref{tbl:tdcf_cqccgmm} and \ref{tbl:tdcf_lcnn} show the t-DCF values for the combined CQCC-GMM CM and GMM-UBM-based ASV scores and for the combined LCNN CM and GMM-UBM-based ASV scores, respectively.
Compared to the conventional playback attacks, an attack using our enhanced stolen speech method greatly degraded the authentication performance of both combinations.
This suggests that our enhanced stolen speech method enables playback attacks to pass playback spoofing CMs and to fool ASV systems as well.

\vspace{-2mm}
\section{Conclusion and future work}
\label{sec:conclusion}
\vspace{-2mm}
We investigated the effectiveness of using enhanced stolen speech in playback spoofing attacks. Experimental results showed that stolen speech enhanced with SEGAN can greatly degrade the performances of baseline CQCC-GMM and advanced LCNN-based playback spoofing CMs as well as that of GMM-UBM-based ASV systems.

Since the used speech enhancement method for attack would be unknown, we plan to develop a robust playback detection method for various speech enhancement methods.

\section*{Acknowledgement}
This work was partially supported by JSPS KAKENHI Grant Numbers JP16H06302, 18H04120, 18H04112, 18KT0051, 17H04687, and Academy of Finland (project no. 309629). We thank Huy H. Nguyen, the Graduate University for Advanced Studies (SOKENDAI), for comments on an earlier versions of the manuscript.

\bibliographystyle{ieee}
\bibliography{refs}

\end{document}